\documentclass[runningheads]{llncs}

\usepackage[nolist]{acronym}
\usepackage{comment}
\newacro{vr}[VR]{Virtual Reality}
\newacro{hmd}[HMD]{Head-Mounted Display}
\newacro{ems}[EMS]{Electrical Muscle Stimulation}
\newacro{imu}[IMU]{Inertial Measurement Unit}
\newacro{bci}[BCI]{Brain-Computer Interface}
\newacro{vwg}[VWG]{Virtual World Generator}
\newacro{moo}[MOO]{Multi-Objective Optimization}
\newacro{dof}[DoF]{Degree of Freedom}
\newacro{ssq}[SSQ]{Simulator Sickness Questionnaire}
\newacro{fov}[FOV]{Field of View}
\newacro{plt}[PLT]{Pareto Least Turns}
\newacro{psp}[PSP]{Pareto Shortest Path}
\newacro{rrt}[RRT]{Rapidly-exploring Random Trees}
\newacro{vims}[VIMS]{Visually Induced Motion Sickness}
\newacro{ddr}[DDR]{Differential Drive Robot}
\newacro{vrise}[VRISE]{VR Induced Symptoms and Effects}
\usepackage{graphicx}
\usepackage{amsmath} % assumes amsmath package installed
\usepackage{amssymb}  % assumes amsmath package installed
\usepackage{verbatim}
\usepackage{lmodern,textcomp}
\usepackage{gensymb}

\begin{document}

\title{Comfort and Sickness while Virtually Aboard
an Autonomous Telepresence Robot %\thanks{Supported by organization x.}}
}
\titlerunning{Comfort Aboard a Telepresence Robot}
\author{Markku Suomalainen\inst{1} \and
Katherine J. Mimnaugh\inst{1} \and
Israel Becerra\inst{2} \and
Eliezer Lozano\inst{2} \and
Rafael Murrieta-Cid\inst{2} \and
Steven M. LaValle\inst{1}}
\authorrunning{M. Suomalainen et al.}
\institute{Center for Ubiquitous Computing, University of Oulu, Oulu, Finland
\email{firstname.surname@oulu.fi}
\and
 Centro de Investigacion en Matematicas (CIMAT), Guanajuato, Mexico
\email{\{israelb,eliezer.lozano,murrieta\}@cimat.mx}}

\maketitle              % typeset the header of the contribution
%%

\begin{abstract}
  In this paper, we analyze how different path aspects affect a user's experience, mainly VR sickness and overall comfort, while immersed in an autonomously moving telepresence robot through a virtual reality headset. In particular, we focus on how the robot turns and the distance it keeps from objects, with the goal of planning suitable trajectories for an autonomously moving immersive telepresence robot in mind; rotational acceleration is known for causing the majority of VR sickness, and distance to objects modulates the optical flow. We ran a within-subjects user study (n = 36, women = 18) in which the participants watched three panoramic videos recorded in a virtual museum while aboard an autonomously moving telepresence robot taking three different paths varying in aspects such as turns, speeds, or distances to walls and objects. We found a moderate correlation between the users' sickness as measured by the SSQ and comfort on a 6-point Likert scale across all paths. However, we detected no association between sickness and the choice of the most comfortable path, showing that sickness is not the only factor affecting the comfort of the user. The subjective experience of turn speed did not correlate with either the SSQ scores or comfort, even though people often mentioned turning speed as a source of discomfort in the open-ended questions. Through exploring the open-ended answers more carefully, a possible reason is that the length and lack of predictability also play a large role in making people observe turns as uncomfortable. A larger subjective distance from walls and objects increased comfort and decreased sickness both in quantitative and qualitative data.  Finally, the SSQ subscales and total weighted scores showed differences by age group and by gender. 
  

\end{abstract}

\section{Introduction}\label{sec:intro}

\begin{figure}[t]
  \includegraphics[width=\textwidth]{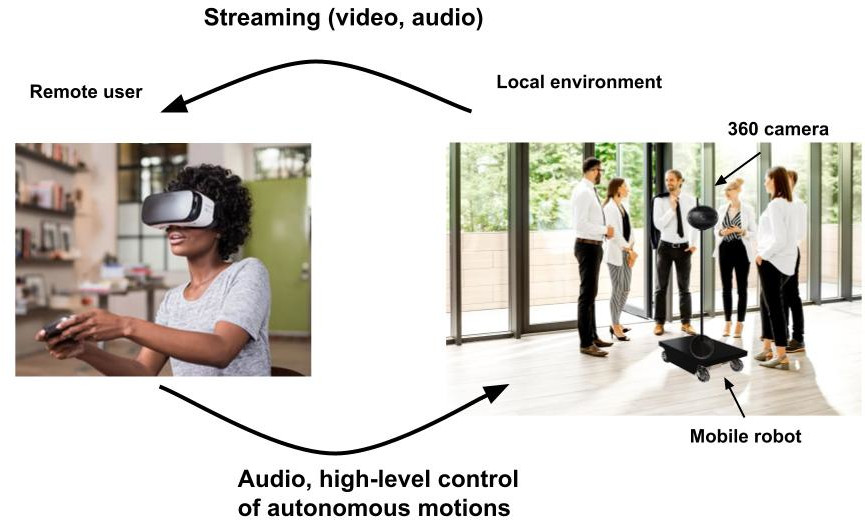}
  \caption{Immersive robotic telepresence.}
  \label{fig:teaser}
\end{figure}

In immersive robotic telepresence, as seen in Fig.~\ref{fig:teaser}, a user wearing a \ac{hmd} embodies a physical robot in a distant location. Besides the visual input through the \ac{hmd}, the user can communicate with people around the robot through bidirectional audio, and can command the robot to move. This technology creates opportunities to visit museums or nature for people with limited mobility or who otherwise cannot, or grandparents attending grandchildren's birthdays far away. Overall, the technology enables meetings which mix physically present and remotely attending people, such that, for example, a single remote participant can join a physical meeting and actually feel as if she was really there.
The robot also allows for touring real facilities or office buildings, helps join physical conferences remotely and facilitates the important impromptu corridor discussions during remote work \cite{tsui2011exploring}.

There is evidence that telepresence robot users prefer the robot to handle low-level motions autonomously, such that the users only choose the target location for the robot to move in a map or within visible area \cite{rae2017robotic,baker2020towards}. Most of current commercial telepresence robots (with standard camera streaming into a standard screen, in other words, not employing an \ac{hmd}) have a joystick or similar as the main control method, with autonomy functionalities only being rolled out at the moment in products such as the Double 3 robot. However, with the increased immersion and embodiment of an \ac{hmd}-based telepresence robot, any autonomous path is not suitable but the motions the robot makes must be carefully planned to avoid adversarial effects; pure \ac{vr} research and applications usually enable and encourage the use of teleportation to avoid VR sickness, often caused by sensory mismatch rising from seeing motion in the \ac{hmd} while staying stationary. Thus, there is a very limited amount of research on how an immersive telepresence robot should move to make the embodied user feel comfortable.

%which is infeasible in telepresence; thus, autonomous continuous paths have been investigated mainly in environments not traversable by a human, \cite{}, such as paths through a virtual version of a human for medical use cases \cite{al2018virtual,mikula20113d,mirhosseini2017automatic}. 

%Moreover, teleportation causes more disorientation than continuous motions \textcolor{red}{Find ref}, which can cause anxiety and discomfort \cite{darken2014spatial}. Besides being an important piece in telepresence, should continuous autonomous motions be researched more for virtual worlds as well \textcolor{red}{need to properly look this up if we want to make this proposal}. Also, sometimes teleport makes people more sick than continuous control \cite{clifton2020effects}. 

It is unclear which aspects of autonomous motions make an \ac{hmd} user feel uncomfortable or experience VR sickness \cite{LaValle_bookVR,Chang_Kim_Yoo_2020}, besides the known result of rotations causing more sickness than translations \cite{keshavarz2011axis}. For example, distance to walls and objects modulates the optical flow linked to sickness, but simply staying as far as possible from objects and walls is often not a suitable path planning strategy for a robot in complex environments. Also, even though there is evidence that performing turns faster may reduce cybersickness \cite{widdowson2019assessing,becerra2020human}, such fast motions may not feel comfortable for the immersed users and there may be more factors in turns besides the speed that modulate comfort and sickness. 

%How is the VR sickness experienced under autonomous navigation, in comparison to traditional VR experiences where the user is in control? 
%Does VR sickness manifest in a different cluster of symptoms or on a different time course than those that result during a typical VR experience where the user is in control? 
%Besides the earlier mentioned aspects,
%good subjective spatial awareness can be important, as lack of it can make users feel anxiety and discontent \cite{darken2014spatial}. With this paper, we want to explore the aforementioned questions to pave way for future with next-generation \ac{hmd}-based telepresence.

%There are several other aspects that can affect the comfort of the user and VR sickness, such as the speed of translation \cite{So_Lo_Ho_2001}, rotation \cite{hu1999systematic}., and the distance to walls and objects. These attributes are not often studied in pure VR research either, since motions in VR are usually either directly controlled by the user, or teleportation is used, which may be more comfortable but is infeasible in telepresence. 

%A possible reason for this is the shown increase in sickness when the user is not in control \textcolor{red}{Is there such a thing?}, even though there's also evidence that teleport can be more sickening for some users . However, Becerra et al.~\cite{becerra2020human} showed that it is possible to reduce the VR sickness even when the user is moving without control. \textcolor{red}{Hmm, this actualy sidetracks the story...}

In this paper we make an attempt to disentangle comfort and VR sickness experienced by subjects aboard an autonomous immersive telepresence robot and advise what aspects in the trajectory of such a robot should be paid attention to. First, %we detected a moderate, negative correlation between VR sickness and the perceived comfort of each path; however, there was no relationship between the user's highest SSQ scores and their choice of the most comfortable path, 
we present an unexpected result: the amount of VR sickness suffered by users does not affect their choice of most comfortable path, even though paths are shown to induce different amounts of VR sickness. As there is a correlation between experienced VR sickness and comfort, it was expected that people who suffer more from sickness would prefer paths that are more comfortable. Then, we present several results regarding the robot's turns; the Likert-scale answers show that perceived turn speed does not correlate with VR sickness or perceived comfort. However, turns are very often mentioned in open-ended questions as reason for discomfort or the choice of most comfortable path; further analysis of the open-ended questions reveals that even though turn speed was the most mentioned keyword, predictability and length of turn also play major roles in making a turn comfortable or uncomfortable. Additionally, distances to walls and objects were weakly correlated with comfort and sickness, which was also reflected in the open-ended questions, meaning that distance should be kept whenever possible. Finally, we report the effect of age, gender and gaming experience on the results, as well as the \ac{ssq} subscales, finding that women in our study experienced higher levels of sickness on average than men, adding to the currently conflicting results on the topic  \cite{Grassini_Laumann_2020,Peck_Sockol_Hancock_2020}. Contrary to a recent meta-analysis \cite{saredakis2020factors}, in this study older subjects suffered more from the effects of VR sickness. We discuss why subjects in our experiments suffered more from VR sickness when compared to roller coaster studies and find ways to reduce the VR sickness caused by VR-based telepresence to make it a viable option for the general population. 

The main contributions of this paper are 1) quantitative analyses of the relation between sickness, comfort, and other variables related to immersive telepresence, with suggestions on turn lengths and speeds, and 2) in-depth analysis of open-ended questions regarding turns, leading to findings that turn predictability and length play major roles in making turns comfortable while retaining the non-sickening abilities of performing the turns fast. Additionally, we present results on VR sickness; carryover effects, effect of demographics and comparison to similar studies, to contribute to the literature on the topic from the perspective of autonomous motions in VR. We note that we performed the experiments purely in VR to avoid confounding factors (such as shaking of the robot) that would arise from using a physical robot. However, we are planning to test the results also on a physical robot to confirm the results.

%The data used in this paper has already been analyzed and used in [redacted for blind review], where we established that piecewise linear paths reduced cybersickness and....

%However, simply attempting to alleviate only a few causes of VR sickness alone is not sufficient to make the use of a telepresence robot a positive experience for the user. 

\section{Related work}
Telepresence, a term originally coined by Minsky \cite{minsky1980telepresence} and sometimes referred to as tele-embodiment \cite{paulos2001social}, is classically, and in this paper, defined as embodying a robot in a remote location. %This location is possibly far from the user (even though nowadays the word can be used with, for example, video conference calls \cite{tang2012social}).
Most of the work on robotic telepresence considers seeing the remote environment through a standard naked-eye display \cite{lee2011now,rae2014bodies,rae2017robotic,neustaedter2018being,fitter2020we}; these works present the potential of telepresence robots in, for example, conferences and classrooms, with also medical applications being an often researched topic \cite{hilty2020review,soares2017mobile}. Additionally, there is research for more personal, intimate, and extensible use cases, such as sharing outdoor activities \cite{heshmat2018geocaching} or more personal long-distance relationships \cite{yang2018our}.

There is an increasing number of works that demonstrate the potential of increased immmersion created by an \ac{hmd} in telepresence, such as better task performance \cite{garcia2015natural} and situational awareness \cite{martins2009immersive}. An interesting motivator is also the finding that in a group work task, users telepresent through a traditional display speak less and perceive tasks as more difficult than the physical participants \cite{stoll2018wait}; this is exactly the sort of issue that
the increased feeling of presence by the user \cite{sanchez2005presence}, facilitated by the increased immersion of an \ac{hmd}, can remedy. A few other researchers have also noticed the importance of robotic telepresence using an \ac{hmd}: Baker et al.~\cite{baker2020towards} let the user choose the destination similarly as in teleporting and then make the robot move to the destination autonomously. Zhang et al.~\cite{zhang2018detection} explored using redirected walking on a telepresence robot. Finally, Oh et al.~\cite{oh2018360} used such a robot on a tour with several pre-selected destinations.

%When limited to continuous motions in telepresence, the path needs to be planned such that the advantages are exploited while the disadvantages are minimized.
%There are only a few works that consider continuous autonomous motions in a human-traversable environment,
%For example, there is evidence that continuous motions increase the feeling of presence over time and help retain wayfinding ability when compared to teleportation \cite{clifton2020effects,moghadam2018scene}. 
%Baker et al.~\cite{baker2020towards} presented waypoint navigation, a way to choose a target within visible area similarly as teleportation in a virtual environment, after which the robot moves there autonomously; they found that users preferred this over joysticking, confirming the feedback presented in \cite{rae2017robotic}. When comparing manual and autonomous motions, in an early work there was no difference in wayfinding regardless of the amount of control the user had \cite{bowman1999maintaining}. However, at present, the authors are not aware of research on whether different kinds of autonomous paths help the users to better retain their wayfinding ability. 

However, as always when using an \ac{hmd}, VR sickness %, also known as \ac{vrise}, 
can severely deteriorate the experience for many users. A major cause for VR sickness is a conflict between the visual and vestibular senses when self-motion is seen through the \ac{hmd} but not sensed by the vestibular organs in the ears \cite{reason1975motion,Chang_Kim_Yoo_2020}. In virtual environments, the use of teleportation is frequently used to avoid VR sickness \cite{Buttussi_Chittaro_2020}, but in telepresence teleportation would have a significant delay and is thus not a realistic option. %Additionally, continuous motions offer  (especially since some people actually get more sick using teleportation \cite{clifton2020effects}). 
Thus, in virtual environments, continuous autonomous motions are often avoided, with perhaps the notable exception of roller coasters often used in sickness studies due to their more sickening effects \cite{davis2015comparing,nesbitt2017correlating,mchugh2019investigating,islam2020automatic}. Though there is a recommendation that VR sessions should not last longer than 55-70 minutes to avoid overwhelming sickness levels \cite{kourtesis2019validation}, in the roller coaster studies many people could not even complete a 15 minute session. A meta-analysis found that 15.6\% of participants across 46 VR studies dropped out due to sickness effects \cite{saredakis2020factors}.

Even though sickness plays a major part in the comfort of an embodied telepresence user, it should not be the only criterion considered when designing motions for a telepresence robot. Becerra et al.~\cite{becerra2020human} showed that the use of piecewise linear paths (meaning that the robot is only rotating in place, and during forward motions there was no rotation) can decrease the VR sickness on embodied participants and make the path more comfortable when compared to a traditional robot path where the robot can rotate and move forward simultaneously; however, they did not explore more detailed questions on why such turns were preferred, or other path aspects such as speeds and distances to objects. Moreover, due to the varying susceptibility to VR sickness \cite{rebenitsch2014individual}, other aspects should not be completely ignored in favor of reducing VR sickness, as some methods to reduce VR sickness (such as driving extremely slowly) also deteriorate the overall experience. To the knowledge of the authors, there are no studies that would consider aspects such as closeness to objects and speeds as factor in immersive telepresence robots motion planning. 
%Thus, it is important to observe how different preferences affect the choice of path. Becerra et al.~\cite{becerra2020human} introduced human perception-optimized planning, which used Pareto-optimization as the tool to study the interplay between various preferences (weightings) between optimization criteria. They presented a method to compute the Pareto front, meaning the set of paths where one criterion cannot be improved without deteriorating another criterion. We employ this method to generate paths optimizing various criteria, compare to a baseline method, and observe the relationships between sickness, perceived comfort and wayfinding. 

To collect results from VR studies that give a more accurate reflection of effects in the general population, demographic information and individual differences must be taken into account. Though there have been conflicting findings in the VR literature regarding gender differences in response to VR sickness \cite{Grassini_Laumann_2020}, a recent meta-analysis found that SSQ effects were systematically associated with the number of men and women in the study \cite{Peck_Sockol_Hancock_2020}, such that when there are fewer women participants, SSQ scores tend to be higher. In regards to age differences in VR sickness effects as measured by the \ac{ssq}, a 2020 meta-analysis of 55 VR articles analyzing VR sickness across types of content and individual factors found that older subjects had lower SSQ scores, and in particular older subjects had significantly lower SSQ scores on the disorientation subscale as compared with younger subjects \cite{saredakis2020factors}.

\section{Methods}

\subsection{Setup and test paths}
\begin{figure}
\centering
\includegraphics[width=\columnwidth]{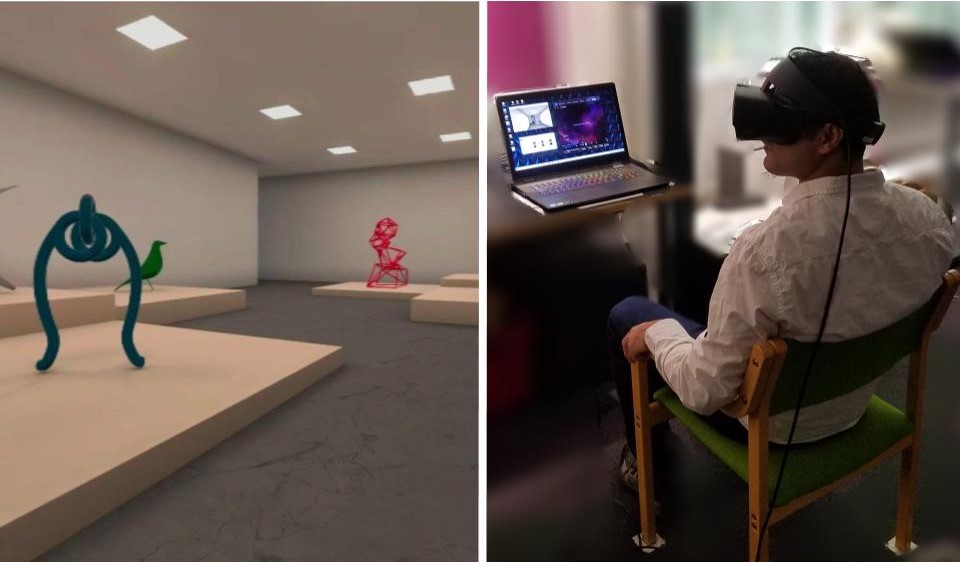}
\caption{A participant in the user study watches one of the panoramic videos in the Oculus Rift S (right). A screenshot of the gallery inside the virtual museum (left).}  %The videos used and other relevant documents are available at: \textcolor{red}{[redacted for blind review]}
\label{fig:teleop} 
\vspace{-0.2cm}
\end{figure}

The study was run on a university campus in February 2020 before COVID-19 caused any local restrictions. The users were seated during the study, as shown in Fig.~\ref{fig:teleop}, and shown pre-recorded panoramic videos with an Oculus Rift S using the Virtual Desktop application; as the video was recorded as a 360 video, the users were able to look around in the virtual environment as the virtual robot was moving. The test environment was designed with Unity. To create a more realistic optical flow and experience for the users, there were various paintings on the walls and statues in the gallery of the museum. 

\begin{figure}[t]
    \centering
    \includegraphics[width=\columnwidth]{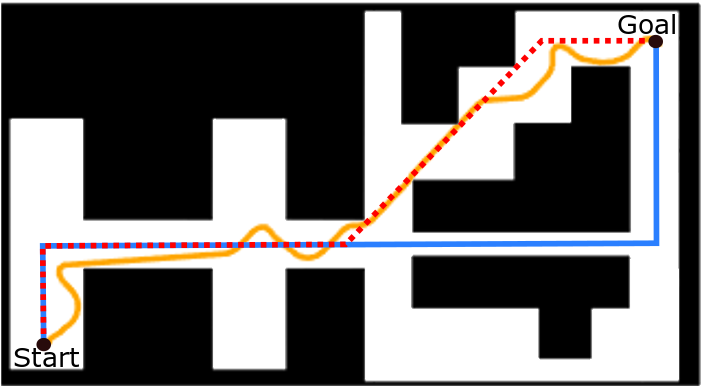}
    \caption{The paths presented to the subjects in the user study: blue is the Pareto Least Turns (PLT), dotted red the Pareto Shortest Path (PSP) and yellow the RRT path.}
    \label{fig:trajectories} 
    \vspace{-3.5mm}
\end{figure}

Three paths were considered, presented in Fig.~\ref{fig:trajectories}. Two of the paths, \ac{plt} and \ac{psp}, were piecewise linear, and in \cite{becerra2020human} they were shown to be less sickening and more comfortable than the third path, the \ac{rrt} path. PLT and PSP were paths chosen from a Pareto front across multiple criteria; the Pareto front of a multiobjective optimization problem is the set of solutions from which any objective score cannot be improved without deteriorating another objective. The Pareto front was computed with the objectives of minimizing the number of turns, minimizing the distance to goal and minimizing the amount of time an object was closer than 2m from the subject. 

The first path, the \ac{plt}, (duration 76s, length 72m, min.~distance to walls 1m), minimized the number of turns along a piecewise linear path (2 turns), and the second path, the \ac{psp}, (67s, 62.6m, 0.4m), minimized the length of the path (4 turns). Both the PLT and PSP had a constant turn speed of 90 deg/s and mean forward speed of 1m/s. The RRT path had smooth turns and was generated by the \ac{rrt} algorithm \cite{lavalle2001randomized}, widely used in robot motion planning. This algorithm respects the dynamics of the \ac{ddr} base, which means that the actual path exists in a 5-D space, from which only two dimensions are plotted here; this is why simply curating the path to avoid unnecessary curves, or finding a more optimal path, is infeasible. As the \ac{rrt} is a sampling-based algorithm and does not provide an optimal path, the algorithm was run 1000 times and the path with the least amount of changes in direction was chosen as the third path (129s, 67.2m, 0.3m, avg.~turn speed 18.8 deg/s, avg.~forward speed 0.52m/s). 

\subsection{Participants}
Subjects were recruited from a university campus and the surrounding community. Of the 45 participants, altogether nine subjects were excluded; three quit the experiment due to excessive sickness symptoms (we will discuss possible reasons for this high number of dropouts in Section~\ref{sec:discussion}), and the rest were excluded due to technical failures, not completing questionnaires, or for feeling severe sickness symptoms before the study began. Thus, the results are from 36 subjects, age 20-44 with mean 28.25, divided equally between men (n = 18) and women (n = 18). Regarding video game use, 25$\%$ of the subjects (n = 9) reported that they never played video games on PC, mobile, or console, 14$\%$ (n = 5) rarely played them, 28$\%$ (n = 10) played them weekly or often, 19$\%$ (n = 7) played them daily, and 14$\%$ (n = 5) did not respond.

\subsection{Measures}
The Simulator Sickness Questionnaire (SSQ) \cite{kennedy1993simulator} was used as a measure of VR sickness symptoms. Before seeing any of the path videos, subjects completed a baseline SSQ which was used for screening purposes. They also filled out SSQ's immediately after seeing each video; the order in which the videos were shown to users was counterbalanced to avoid order effects. Then, the subjects completed a questionnaire created for this study after each path. The questionnaire consisted of several questions that were rated on a six-point Likert scale, followed by some open-ended questions asking why they gave that rating. The Likert-rated questions asked subjects how comfortable their experience was on the path (from very uncomfortable to very comfortable), how well they could find their way back to where they started in the museum (from not easily at all to very easily), how they felt about the distance between themselves and the walls or objects in the museum (from too close to too far), and how they felt about the speed of the turns (from too slow to too fast). %They also rated the forward speed and naturalness of each path. 
The open-ended questions asked why a path was uncomfortable if they had given that rating, and if they had any comments. After they had seen all three paths, they answered additional questions asking them to select which of the paths was preferred and most comfortable, and the reason for that choice.

\subsection{Procedure}
Upon arrival, each subject was given an information sheet about the study and asked to sign a consent form if they wanted to participate. A baseline \ac{ssq} was administered, and then the subject was seated in the experiment chair (Fig.~\ref{fig:teleop}). The experimenter read the instructions out loud for the subjects, and then the subject put on the \ac{hmd} and the first path video was played. After the video, the subject removed the \ac{hmd}, filled in an \ac{ssq}, answered the Likert scale and open-ended questions and drew the path. This process was not timed, but the duration between videos was approximately five minutes. Then, the second video was played and the procedure was repeated. After the third video, the subject completed the final questionnaire comparing the paths, and then reported demographic information and gaming experience. Once they had completed this, they were given a debriefing about the study, copies of the consent forms to take home, and a coupon worth {€}2 for a coffee from the local cafe.

\section{Results}
Data from the same experiment has earlier been used in \cite{becerra2020human,mimnaugh2021analysis}, %where the data in Fig.~\ref{fig:path_main_attributes} and Fig.~\ref{fig:ssq_per_path} have been reported, 
where the focus has been on both the technical implementation, comparison of the paths and the naturalness and preference of the user. In contrast, in this paper we do not focus on comparing the paths and path-planning methods, but instead analyze other interesting results accross all paths. The main focus is about perceived user comfort, to present the community more information about the use of autonomous motions in VR-based telepresence. The total number of comments on turns and distances for the "why was the path uncomfortable" question have been reported in \cite{mimnaugh2021analysis}, but without the more detailed analysis presented in this paper regarding the actual contents of the open-ended comments. Also, even though individual results on questions regarding comfort across paths have been reported, the relationship analyses between variables reported here have not been presented. Finally, we note that \cite{becerra2020human} found PSP and PLT causing less VR sickness and being rated more comfortable than the RRT.

Exploratory analyses were conducted with two-tailed significance levels for alpha set at 0.05 and confidence intervals set at 95\%. When multiple tests were run, Bonferroni correction within each test was used. Post-hoc power analyses and observed effect sizes \cite{Cohen_1988} %(Kendall's W, Cohen's w, and $\eta^2$) 
were calculated using G*Power \cite{Faul_Erdfelder_Lang_Buchner_2007}, Psychometrica freeware \cite{lenhard2016calculation}, or by hand \cite{Tomczak2014}. Thematic analysis with an inductive approach \cite{Patton2005qualitative} was used to classify the responses to open-ended questions by two independent coders. Results from Likert-scale questions, forced-choice questions and SSQ are first presented in Section~\ref{sec:quant}, after which responses to the open-ended questions are analyzed in Section~\ref{sec:qual}.

%\subsection{Sickness}
%\katherine{Will add section for OE answers}

\subsection{Quantitative data}
\label{sec:quant}

\subsubsection{Sickness had a negative correlation with comfort.} The more the subjects suffered from VR sickness, the less comfortable they felt. A Spearman's rank-order correlation test was run between the Likert comfort ratings, from very uncomfortable (1) to very comfortable (6), and the total weighted SSQ scores after each path. There was a statistically significant moderate, negative correlation $(rs(108) = -.400, p < .001)$; as the total SSQ scores increased, the Likert comfort ratings decreased. 

\subsubsection{High sickness scores did not influence people's choice of preferred or most comfortable path.} The motivation for checking for this relationship is the assumption that people who suffer more from VR sickness would put a higher weight on how comfortable the path is. To crudely rank the sickness sensitivity of the participants, the highest of the three total SSQ scores for each individual was selected as an index. These scores were then separated into three equal-sized groups of 12 people. Highest total weighted SSQ scores under 15 were in the low sickness group, between 15 and 40 in the medium sickness group, and over 40 in the high sickness group. The relationship between sickness groups and choice of most comfortable path was analyzed using crosstabulation and a Fisher's exact test (two-sided) to account for the small sample size. There was no statistically significant association in choice of the most comfortable path by sickness group, $p = .164, w = 0.85$. Similarly, there was no statistically significant association between choice of preferred path and sickness group, $p = .873, w = 0.58$.

\subsubsection{Increase in perceived distance to walls and objects had a weak correlation with sickness and comfort.} The distance to walls and objects in virtual museum from 1 (too close) to 6 (too far) was compared to Likert ratings for comfort (higher numbers mean greater comfort) and total weighted SSQ scores (higher scores mean more sickness symptoms) using a Spearman's rank-order correlation test. There was a statistically significant weak, positive correlation $(rs(108) = .302, p = .002)$ between the distance to walls and objects and comfort. The closer to walls and objects the paths were rated, the less comfort subjects reported experiencing. Compared to the SSQ scores, there was again a statistically significant weak, negative correlation $(rs(108) = -.277, p = .004)$. The closer to walls and objects the paths were rated, the more sickness symptoms subjects reported experiencing.

\subsubsection{The perceived speed of turns did not influence comfort or sickness} There was no statistically significant correlation between the speed of the turns, rated from too slow (1) to too fast (6), and the Likert ratings of comfort $(rs(108) = -.157, p = .105)$ or the total weighted SSQ scores $(rs(108) = .117, p = .228)$.

\subsubsection{There were no carryover effects on sickness with 5 minute breaks between videos.} The videos of the museum were counterbalanced by gender and by the order that they were seen in. %This counterbalancing was used to counteract any carryover effects from one video impacting sickness scores on subsequent videos. To test if the counterbalancing was successful, 
This counterbalancing was used to allow comparison between paths regardless of whether carryover effects had an impact or not. However, to see if there were carryover effects, a Friedman's test on the SSQ total weighted scores after each video by order was run. There was no statistically significant difference in the distributions of total weighted SSQ scores after the first, second, and third videos, $\chi^2(2,36) = .775, p = .679, W = 0.01$. %Despite subjects reporting high levels of VR sickness symptoms
Thus, the recovery time subjects had with the headset off, involving answering questionnaires for about five minutes after each video, appears to have been sufficient for participants to recover from these VR sickness effects. In one extreme case, for example, the subject's total weighted SSQ score after their second video was 153.34, and the score after their third video was zero. Whereas several papers report longer carryover effects (for example \cite{singer1998virtual}), a large variation across studies has been observed, with lowest recovery times being 10 minutes \cite{duzmanska2018can}.

\subsubsection{Older people suffered more from VR sickness.} 
\begin{figure}[t]
    \centering
    \includegraphics[width=\columnwidth]{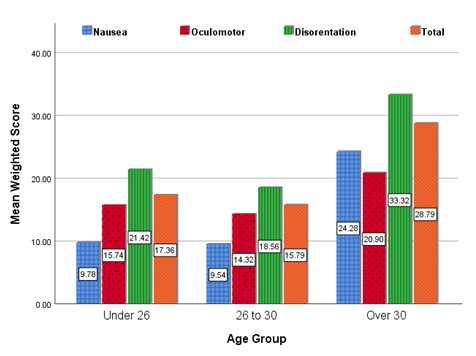} 
    \caption{Mean  weighted SSQ subscales and total scores from all paths by age group.}
    \label{tab:ssq_age_graph} 
    \vspace{-3.5mm}
\end{figure}

Age differences in sickness were examined by splitting the subjects into three similar-sized groups for analysis (small variance in group sizes result from not splitting same-aged subjects to different groups): Under 26 (13 subjects), 26 to 30 (12 subjects), and Over 30 (11 subjects), collapsing across paths, and testing the difference between these groups on each of the SSQ subscales and total weighted scores. A Kruskal-Wallis test showed for the nausea subscale (NS), $\chi^2(2,108) = 19.02, p < 0.001, \eta^2 = 0.16$, the disorientation subscale (DS), $\chi^2(2,108) = 9.23, p = 0.010, \eta^2 = 0.07$, and the total weighted score (TS), $\chi^2(2,108) = 11.48, p = 0.003, \eta^2 = 0.09$, there were statistically significant differences in scores between the age groups. Post-hoc pairwise comparisons all showed the same pattern, with the Over 30 age group significantly higher than the 26 to 30 group (NS $p = .001$, DS $p = .019$, TS $p = .012$), and the Over 30 group significantly higher than the Under 26 group (NS $p < .001$, DS $p = .030$, TS $p = .007$), but no difference between the Under 26 group and the 26 to 30 group. The exception to this pattern was on the oculomotor subscale, where there was no statistically significant difference between the age groups, $\chi^2(2,108) = 4.296, p = 0.117, \eta^2 = 0.02$. Mean SSQ subscales and total scores for each age group are shown in Fig. \ref{tab:ssq_age_graph}.

\subsubsection{Previous VR experience did not influence VR sickness.}  Subjects were split into three groups based on how often they used virtual reality to test whether or not previous virtual reality experience had an impact on sickness. Ten subjects had never tried a VR HMD (28\% of the sample), 15 subjects had tried VR between one and nine times or a few times (43\% of the sample), and 10 subjects tried ten or more times or used VR regularly (28\% of the sample). One subject did not give a response about his previous VR use and was excluded from these analyses. There were no statistically significant correlations between VR usage frequency and the nausea subscale $(rs(105) = .011, p = .915)$, the oculomotor subscale $(rs(105) = -.030, p = .764)$, the disorientation subscale $(rs(105) = .038, p = .697)$, or the total weighted SSQ score $(rs(105) = .008, p = .933)$.

\subsubsection{Women suffered more from VR sickness.} 
\begin{figure}[th!]
    \centering
    \includegraphics[width=\columnwidth]{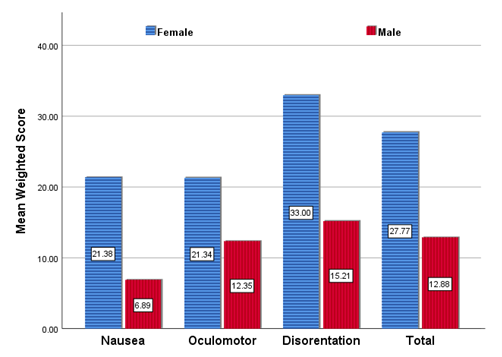} 
    \caption{Mean weighted SSQ subscales and total scores from all paths by gender.}
    \label{tab:ssq_gender_graph} 
    \vspace{-3mm}
\end{figure}

To investigate potential differences in sickness by gender, scores across paths were collapsed and compared each with a Mann-Whitney U test. Women had statistically significantly higher SSQ scores than men on the nausea subscale $(U = 746.00, z = -4.55, p < .001, r = 0.44)$, the oculomotor subscale $(U = 1037.50, z = -2.63, p = .008, r = 0.25)$, the disorientation subscale $(U = 904.50, z = -3.51, p < .001, r = 0.34)$, and the total weighted score $(U = 827.50, z = -3.90, p < .001, r = 0.38)$, as presented in Fig. \ref{tab:ssq_gender_graph}.

\subsubsection{Subjective wayfinding ability was weakly correlated with comfort.} The idea for testing this correlation is reports that loss of wayfinding ability can cause discomfort \cite{darken2014spatial}. Likert ratings for comfort (higher numbers mean greater comfort) were also tested against the Likert ratings for subjective wayfinding (answer to the question "If you had to go back to where you started in the museum, how well do you think you could find your way back?", with a high number meaning more likelihood of finding the way back) using a Spearman's rank-order correlation test. There was a statistically significant weak, positive correlation $(rs(108) = .205, p = .034)$. The better that subjects believed they could find their way back to the beginning of the museum, the greater the comfort experienced.

\subsection{Qualitative data}
\label{sec:qual}

\subsubsection{Turns was the most commented aspect making a path uncomfortable, with fast turns being the most commented within turns but unexpectedness and length of turns also playing major roles.} After the Likert-rated question regarding how comfortable each path was, subjects were asked an open-ended question, "if it [the path] was uncomfortable, why?" Their answers were coded with keywords based on the text that they provided. A full breakdown of the distribution of all codes for each path can be seen in Fig.~\ref{fig:OE_uncomfortable}. The largest group of comments was in relation to the turns, with 59 comments across all paths; fast turns had the greatest number of individual comments, with surprising turns, sharp turns and many turns also getting at least several mentions. This is in contrast to turn speed not being correlated with the Likert-scale value for comfort. 

\begin{figure}[]
    \centering
    \includegraphics[width=\columnwidth]{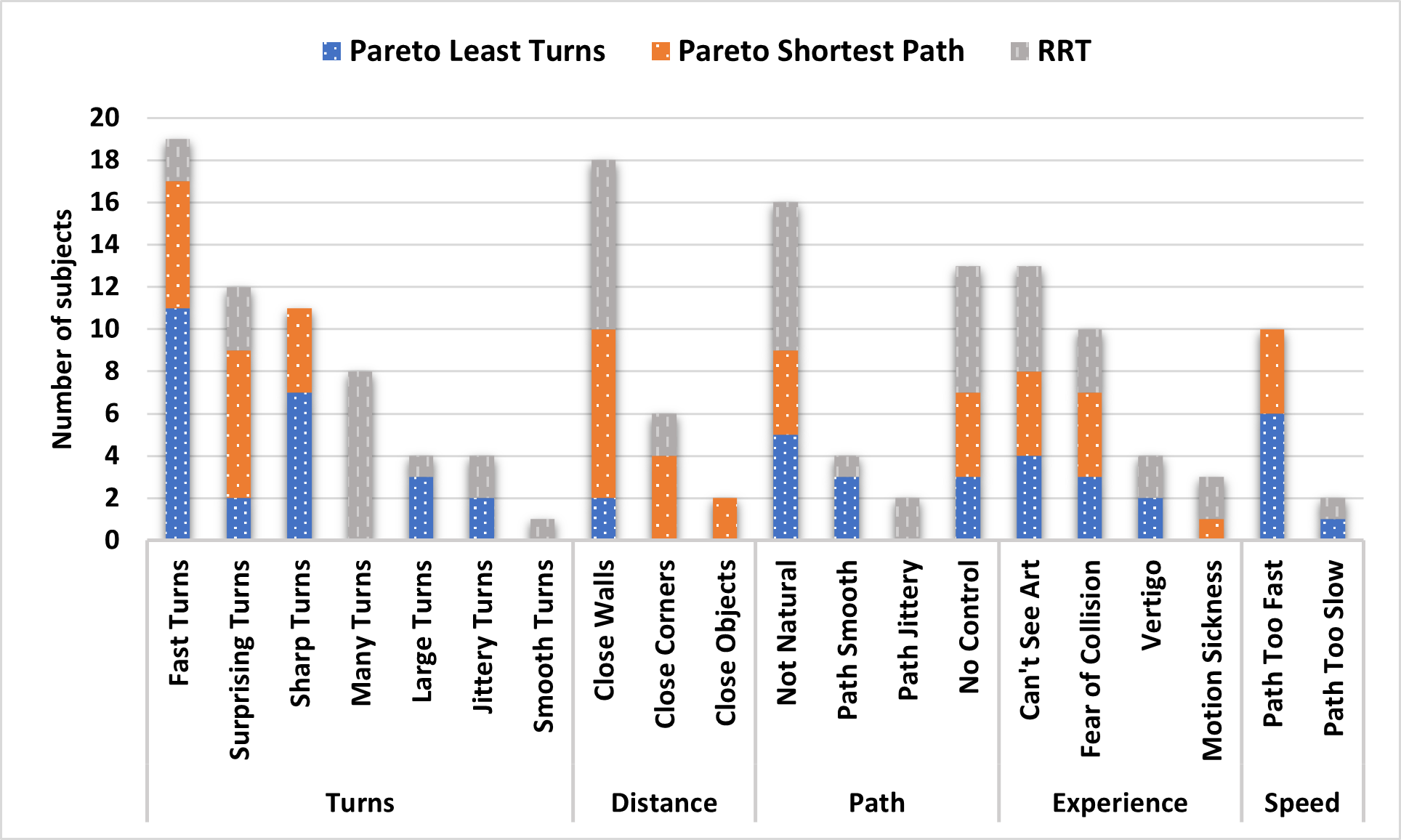} 
    \caption{The frequently found codes from the question "If the path was uncomfortable, why?" which was asked after each path that a each user watched.
    }
    \label{fig:OE_uncomfortable} 
    %\vspace{-3.5mm}
\end{figure}

Whereas fast and surprising turns are quite evident, looking closely into the comments saying "sharp turns" reveals that the word "sharp" is often associated with the "length" of the turn. For example, the comments \textit{"SHARP (90 DEGREE)TURNS ARE VERY UN REALISTIC" } and  \textit{"The turn is very sharp and rapid. Turn is like 90 degree, so I feel uncomfortable"} were typical. Moreover, the degrees of the turn were mentioned also when turns were classified as "fast", such as \textit{"The 90 degrees turns felt very fast."}. It is likely that because there is no clear and concise everyday word, especially within non-native speakers, to talk about the degrees a turn makes, the "large turns" category is in fact larger than it appears to be in these results.%; this evidence points that 90\textdegree turns are considered uncomfortable. 

\subsubsection{Turns also had the largest influence on the choice of the most comfortable path. The "size", speed and predictability of the turn seem to have a large effect.} The largest number of individual comments were under the code "good turns". However, it become evident that people had varying preference of turns when looking at the responses together with the chosen path. For example, subjects chose the PSP because \textit{"It was smoother, there were not so many forced turns"} and \textit{"The walking speed was not too high, and again, the number of uncomfortable turns was not too high."} and \textit{"Turn is not rapid and sharp. The distance is also OK."}, even though PSP had the same amount of 90\textdegree ~turns as PLT, and additionally two 45\textdegree~turns. This implies that, even though long turns at once should make people less sick, users would still prefer not to have such long turns and 45\textdegree~turns do not bother users as much. Additionally, even though only one subject used the word "predictable", terms such as \textit{"forced"} (above) and \textit{"abrupt"} (\textit{"There were not so much abrupt turns."}) were used on the paths not chosen as the most comfortable, indicating that predictability plays a bigger role in making such turns comfortable than what could be deducted from the coding alone.

%\subsubsection{Several people still chose the paths with fast turns as the most comfortable saying that the turn speed was good.} Besides the large amount of comments regarding turns being too fast, all five comments regarding suitable turn speed were on either PLT or PSP, which both had $90\frac{\degree}{s}$ turn speed (compared to the $18.8\frac{\degree}{s}$ turn speed in the RRT path); thus, there was an alement that made the users perceive one path's turns faster than the others. This is likely related to what was discussed in the previous paragraph about turn radiuses; \textit{"THE ROBOT WAS NOT MOVING OR TURNING VERY FAST AND IT WAS FAR FROM OBJECTS " PLT} and \textit{"Turn is not rapid and sharp. The distance is also OK." PSP} and \textit{"THE SEED OF MOVE IS COMFORTABLE, AND THE DISTANCE OF THE WALL IS COMFORTABLE. THE SPEED OF TURN IS ACCEPTABLE." PLT} and \textit{"It has the least turns and the turning speed wasn't uncomfortable." PLT}

\begin{figure}[]
    \centering
    \includegraphics[width=\columnwidth]{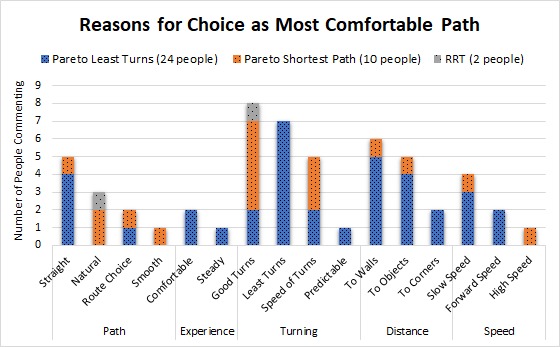} 
    \caption{The frequently found codes from the question "Of the three paths, which one was the most comfortable? Why?" which was asked after the subject had seen all three paths.
    }
    \label{fig:OE_uncomfortable} 
    \vspace{-3.5mm}
\end{figure}

\subsubsection{Small distances to walls and objects were considered a major source of discomfort, and an important factor when choosing the most comfortable path.} There were 26 comments across all paths regarding the distance group of responses in the "why uncomfortable" question, most of which were related to close walls (18) or corners (6), with only two responses mentioning being too close to objects. A very related category is fear of collision (10 mentions), where most quotes were related to distance to walls (\textit{"Sometimes it felt I would bump into the walls"}); these responses were detected in all paths, with PSP and RRT being most mentioned but also PLT gathering several mentions, even though PLT had a larger minimum clearance (1m) than the other two paths (0.3m and 0.4m). The responses reveal that it was considered problematic that the 1m clearance was at a 2m wide corridor, which was deemed too narrow (\textit{"in the end when we went trough a very narrow path, it felt like the walls were closing in."}); indeed, a good distance to objects or walls was often mentioned as a favorable property when choosing the most comfortable path (Fig.~\ref{fig:OE_uncomfortable}), with almost all mentions being together with the PL. When inspecting the comments regarding most comfortable path, the distance to objects is mentioned as often as distance to walls. However, from the five mentions of "distance to objects", only one considered that the objects were close enough (\textit{"going near to the objects"}), whereas another one simply stated that distance was "good" and the rest preferred staying as clear from objects as from walls and corners.

\section{Discussion}
\label{sec:discussion}
%Even though the two Pareto paths weren't significantly different in the Likert rating scores for comfort, the Pareto least turns path was rated by the Likert scores as being almost neutral (mean score 3.72) in the distance to walls and objects, which was significantly different from the the Pareto shortest path (mean 2.64) and the RRT (mean 2.03), which were rated as going too close to walls and objects. 
%Additionally, though the speed of the turns on the Pareto least turns and Pareto shortest path were the same, and there was no significant difference in score on the Likert ratings on turn speed between the two Pareto paths, the Pareto least turns was mentioned as having turns that were too large, too fast, and too sharp more often than the other paths on the open-ended questions. So, even though the turns were less comfortable on the Pareto least turns path, the fact that there were only two of them compared the Pareto shortest path, and the fact that the distance to walls and objects was better, may have contributed to reasons for the Pareto least turns being selected as the most comfortable of the three paths. This is also supported by the fact that the turns were one of the most commented on aspects in relationship to comfort.

We expected to find a stronger link between the perceived comfort and SSQ scores than what the results showed (no relationship between the SSQ and the choice of "most comfortable path," and a moderate correlation between the SSQ and the Likert-scale comfort); we expected people with higher SSQ scores to prefer paths that, in general, caused less sickness (PLT and PSP). The lack of this connection shows that there are more facets to a user's comfort than only the often used SSQ; even though experiencing VR sickness can have the most impact on users, sickness symptoms may not be the most significant part of the VR experience for the users. Several subjects stated this outright, such as \textit{"It was not sickening but not comfortable in the way the (virtual) robot moves"}. The importance of this is highlighted by a strong relationship between perceived comfort and preference reported in \cite{mimnaugh2021analysis}, meaning that people can like things that make them a little sick. Also, comfort is a more subjective feeling than sickness, and discomfort does not necessarily equal sickness; all of this should be taken into account when designing VR experiences and new methods for quantifying the sickness effects and comfort of VR exposure.

%a close distance to objects and walls would have an effect on the SSQ due to an increase in optical flow, and

We were also expecting that perceiving a fast turn speed would not have an effect on the SSQ, based on \cite{widdowson2019assessing} finding that performing the same turn faster than slower makes people less sick; the data confirmed this. However, it came as a surprise that even though the turns were most frequently mentioned in the open-ended questions regarding why users felt a particular path was uncomfortable, there was no statistically significant correlation between Likert-scale turn speed ratings and comfort. A likely reason is that we specifically asked about the turn speeds in the Likert-scale question; thus, the Likert-scale question did not consider any other aspects of the turn besides speed. Besides the earlier mentioned quotes where the 90\textdegree~turns were specifically mentioned, also other quotes with fast turns had additional adjectives hinting towards the surprise element, such as \textit{"the movement was slow but, the turns were fast a jarring"}. Another subject, who chose PSP as the most comfortable path, stated that PSP \textit{"Didn't contain sharp turns and lot's of unreasonable moves. "}, even though it contained the same amount of 90\textdegree ~turns as PLT, and additionally two 45\textdegree~turns; this suggests that the 45\textdegree~turns were not considered as uncomfortable as the longer turns. These findings indicate that there could be even a contradiction between perceived comfort and experienced VR sickness during turns; long turns are considered uncomfortable, even though they reduce sickness. We believe this should be taken into account when designing the motions of a telepresence robot, and make a more controlled experiment about turn speeds and lengths in VR, where besides sickness also comfort is queried. Even though the results are only preliminary, we suggest avoiding 90 degree turns at once and making the turns more predictable. We still suggest keeping the turns fast, at least close to the 90~deg/s, since there is known evidence of effect on VR sickness but contradictory evidence with comfort. 

Distance to walls and objects had a correlation with both comfort and sickness, even if weak, and was also mentioned frequently in the open-ended questions. This was expected, as a closer passing distance increases the optical flow, which is linked to increase in VR sickness. Additionally, fear of collision was mentioned a few times, which, as a strong feeling causing anxiety, may have a strong effect on comfort and sickness \cite{guna2019influence}. Based on the results, 1m passing distance did not cause users any discomfort unless in a narrow corridor, but such a passing distance would make robot motion planning complicated. However, it would be beneficial to verify the result on a real robot in a real environment, with more variance of passing distance.  

The observed relationship between subjective wayfinding ability and comfort is not surprising, since there is evidence in the literature that loss of wayfinding ability can cause discomfort \cite{darken2014spatial}. The environment used in this study was too simple to test objective wayfinding ability, and the correlation is only weak, with also the possibility that this is only a coincidence; however, the observed connection and evidence in literature indicates that a more focused study would be useful to explore whether the wayfinding ability can be increased by careful planning of the robot's motions. 

The subjects succumbed to VR sickness quickly considering that the videos were not very long (67s-129s) and the robot moved slowly ($0.5 m/s-1 m/s$); however, the subjects also recovered faster than suggested by earlier literature. For example, Kourtesis et al.~\cite{kourtesis2019validation} state that VR sessions can be comfortable up to 55 to 70 minutes, but we experienced three dropouts (two during their first videos, which were RRT and PLT, and one while watching RRT as second video after already finished PLT) with videos lasting less than two minutes. This supports the proposition that continuous motions, without a visual, stationary cue that moves with the user, are a significant contributor to VR sickness. However, the surprisingly fast recovery time (no carryover effects detected with 5 minute breaks, whereas the lowest suggested required recovery time from literature is 10 minutes \cite{duzmanska2018can}) balance this; these results hint that VR sickness with continuous motions raises fast but also decreases fast. However, more focused research on VR sickness for continuous motions would be required to confirm this. 

Comparing to the \ac{ssq} scores caused by roller coasters in two other studies (both using HTC Vive), the results from this study were comparable or slightly higher when adjusted for the duration of the stimuli; Islam et al.~\cite{islam2020automatic} had no dropouts in a 15-minute roller coaster ride with a mean SSQ of 55.45, and McHugh et al.~\cite{mchugh2019investigating} had no dropouts in a 5 minutes 50 seconds exposure with a mean SSQ of 34 (visually approximated from figure), whereas our mean SSQ was 20.3; thus, it seems that the telepresence experience provided a higher variance of sickness to test subjects than a roller coaster, since in our study there were more dropouts but less average sickness. In fact, there are a few considerable differences between telepresence and a roller coaster, both adding and reducing the potential for VR sickness. A roller coaster moves much faster, which increases the optical flow and the probability for sickness, but it has a constant frame of reference (the car) which is known to decrease VR sickness \cite{cao2018visually}. Also, most of the scenery does not move with the speed of the roller coaster if it travels higher than on ground level, reducing the optical flow; a reduction of sickness has been reported when a plane flies higher above the ground \cite{johnson2005introduction}. Additionally, the turns are predictable - the user can always see the track ahead, which has also been shown to reduce VR sickness \cite{kolasinski1995simulator}. We also observed individual comments in open-ended questions on turns being unpredictable (\textit{"the turns were too fast, but thankfully pretty predictable this time."}) and, in regards to why the path was uncomfortable, \textit{"The turns were very strong (sudden)"} and \textit{"It was not sickening but not comfortable in the way the (virtual) robot moves. Sudden turns very close to the corners of the walls."} Thus, future ways to combat VR sickness in autonomous telepresence motions could be visualizing the path to the user so that movements are less unexpected and providing a constant frame of reference, such as visualizing the robot. 

%\markku{Maybe mention that ours moved much slower, so in a sense our data fits pretty well with these other roller coasters. Maybe we also had more VR users, which could push down the average. BUT there was no SSQ relationship to VR gaming. BUT the McHugh paper had a stationary cart though. Also the whole floor is moving, not just the tracks, so there's more optical flow. }

Women experienced greater sickness symptoms during the study than men, which has been found previously in VR research \cite{rebenitsch2014individual}). In this study, previous VR experience did not influence the severity of VR sickness, which is in agreement with \cite{kourtesis2019validation}, even though such an effect through habituation \cite{howarth2008characteristics} has been reported elsewhere \cite{rebenitsch2014individual}. %This could be due to continuous motions not often being the typical stimuli in VR applications.
Regarding higher sickness in older subjects, we were unable to find similar work in the literature, and in fact found conflicting results \cite{saredakis2020factors}. One explanation could perhaps be related to differences in lens flexibility in the eye across the lifespan, making it easier to focus on things very close to your face when you're younger and sometimes resulting in the need for reading glasses later in life. It will be an interesting topic for future research whether there is something specific in continuous motions that causes this discrepancy.  

%the effect size, however, was so strong that this is very unlikely to be a pure coincidence. - Wrong!!! We can't say this without multiple studies with a lot more subs

\subsection{Limitations and future work}
%We did not measure head motions or presence. We did not expect a major difference in presence between the paths in such a simple environment, but when we perform further experiments in a simulated and real university environment, we will see if there is a difference in presence across different paths. It would be interesting to establish in future work whether a link exists between presence and perceived naturalness of the robot's path. 
We established in this study that the design of the robot's turns are very important for increasing comfort and reducing VR sickness. Based on the results, we can provide initial suggestions for designing robot motions, but to draw decisive conclusions more research is needed. For example, it seems that a Likert-scale question about the length of the turns should have been asked, to pinpoint with more accuracy which aspect of the turn makes them so uncomfortable. Moreover, we want to test how big a role predictability plays, to make the experience closer to a slow speed roller coaster. The predictability could be done either explicitly, such as using arrows or a path on the ground, or by modifying the velocity profile of the turns; for example, the common trick used in animation to employ high-order derivatives of speed, such as jerk, could be useful to increase predictability by making the turn start slower but then increase the speed to take advantage of the less-sickening nature of performing turns fast \cite{widdowson2019assessing}.

The different durations of the videos could have had an effect on the SSQ measures. However, as the travelled distance was still similar, this means that the higher speed had less of an impact on the experienced VR sickness. The slower RRT path made some unnecessary turns, but it is a de facto standard robot motion planning algorithm, and therefore we believe this to be a useful comparison. However, a more direct comparison between a path similar to the PSP but with smooth turns could also be useful to pinpoint the effect of rotation in place. We did not measure head motions, but as they have been successfully used in predicting VR sickness \cite{jin2018automatic}, it would have been interesting to test their correlation to VR sickness; ad-hoc observations from the study instructors indicate that an increase in sickness was observable by a decrease in the subjects' head motions.

We intentionally had art in the museum to make the study more realistic, both in terms of ecological validity and optical flow. Besides an expected effect on the subjects' preferences, the art also affected the perceived comfort; users reported in open-ended questions issues such as how comfortably they could see the art (\textit{"Many times I was just facing the wall and couldn't see the displayed items"}). Whereas it is important to acknowledge the effect that the situation and environment have on comfort, it is difficult to quantify. In the future we plan to run another study in an empty museum and compare the results; this will give us a proper, quantifiable metric on the importance of context, at least between the cases of museum and a simple path from A to B. 

The study was run in a virtual environment, even though the ideas are meant for a telepresence robot streaming real video footage. Although there is some evidence that 360$^{\circ}$ videos make people more sick than virtual environments \cite{saredakis2020factors}, there is no single study comparing these two cases on sickness (although it has been shown that teleoperation increases stress over a virtual environment, which in turn can increase discomfort and sickness \cite{khenak2020spatial}); moreover, as there is no reason to believe that the mechanisms causing sickness would work differently in a 360 video than in a virtual environment, we expect our results to extend to real video capture. Nevertheless, we will also run a study with a real telepresence robot to verify these results. 

The number of subjects in the study was small. We tried to account for this by using statistical tests that are more sensitive for small samples when appropriate, and by calculating the effect sizes and then running post-hoc power analyses. One final limitation was the choice of a six point Likert scale rating instead of a standard five or seven point scale. This scale was selected so that subjects would need to make a decision either way on the ratings and could not simply select a neutral choice. In future studies, a more traditional scale will be used.

\section{Conclusion}\label{sec:con}
We presented an analysis focusing on the interplay between and possible causes of comfort and sickness of VR telepresence users with regards to the path taken by the robot. In essence, discomfort and sickness are distinct yet overlapping concepts, and more focus should be put on addressing the comfort of the user instead of simply targeting VR sickness since reducing only sickness may cause other issues that degrade the whole experience. As autonomous motions are rarely considered in virtual worlds, this paper will hopefully spark further interest in research from other investigators both in telepresence and in autonomous continuous motions in virtual worlds.


\bibliographystyle{splncs04}
\bibliography{references}


\end{document}